%% file: main.tex
\documentclass{article}
\usepackage{spconf,amsmath,graphicx}
\usepackage{booktabs}
\usepackage{cite}
\usepackage[usenames,dvipsnames]{color}
\usepackage[shortlabels]{enumitem}
\usepackage{hyperref}
\usepackage{makecell}
\usepackage{multirow}
\usepackage{subfigure}
\usepackage{tablefootnote}
\usepackage{tabularx}
\usepackage{tipa}

\title{Inferring Pitch from Coarse Spectral Features}
\name{Danni Ma$^{1}$ ~Neville Ryant$^{2}$ ~Mark Liberman$^{2}$}
\address{$^{1}$Department of Computer and Information Science, University of Pennsylvania, USA\\
$^{2}$Linguistic Data Consortium, University of Pennsylvania, USA}
\begin{document}
%\ninept
%
\maketitle
\begin{abstract}
Fundamental frequency (F0) has long been treated as the physical definition of ``pitch'' in phonetic analysis. But there have been many demonstrations that F0 is at best an approximation to pitch, both in production and in perception: pitch is not F0, and F0 is not pitch. Changes in the pitch involve many articulatory and acoustic covariates; pitch perception often deviates from what F0 analysis predicts; and in fact, quasi-periodic signals from a single voice source are often incompletely characterized by an attempt to define a single time-varying F0. In this paper, we find strong support for the existence of covariates for pitch in aspects of relatively coarse spectra, in which an overtone series is not available. Thus linear regression can predict the pitch of simple vocalizations, produced by an articulatory synthesizer or by human, from single frames of such coarse spectra. Across speakers, and in more complex vocalizations, our experiments indicate that the covariates are not quite so simple, though apparently still available for more sophisticated modeling. On this basis, we propose that the field needs a better way of thinking about speech pitch, just as celestial mechanics requires us to go beyond Newton's point mass approximations to heavenly bodies. \footnote{The code and data are publicly available at \url{https://github.com/dannima/InferringPitch}}
\end{abstract}

\input{01-introduction}
\input{02-human}
\input{03-experiments}
\input{04-synthetic}
\input{05-discussion}
\input{06-conclusion}

% \section{ACKNOWLEDGMENTS}
% \label{sec:ack}

% Do not include acknowledgments in the initial version of the paper submitted for blind review.
% If a paper is accepted, the final camera-ready version can (and probably should) include acknowledgments. 

\bibliographystyle{IEEEbib}
\bibliography{refs}

\end{document}

%% file: 01-introduction.tex
\section{Introduction}
\label{sec:intro}

\begin{figure*}[!htb]
    % \hspace{-1.65cm}
	\includegraphics[width=\linewidth]{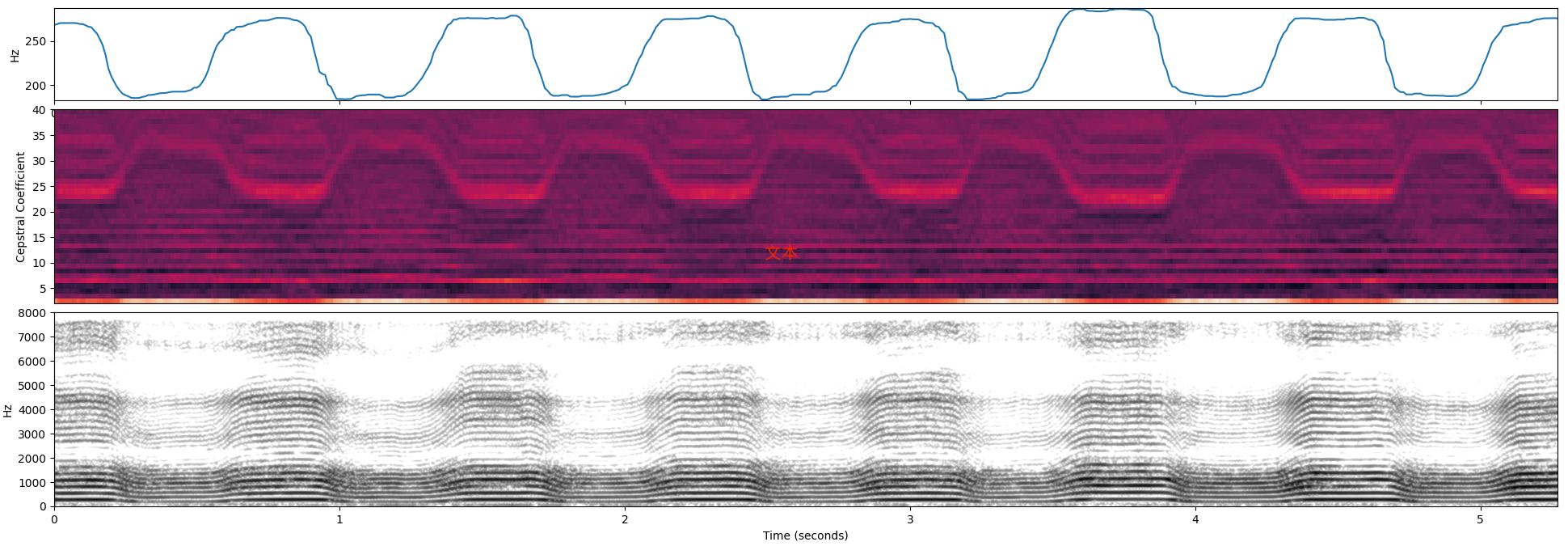}
	\caption{An example production of /\textipa{A}/ from the female speaker. {\bf Top}: F0 estimate from RAPT. {\bf Middle}: Plot of cepstral coefficients (Section~\ref{sec:method} for details). {\bf Bottom}: Log-power spectrum.}
	\label{fig:female_a_eg}
\end{figure*}

Pitch describes how human ears and brains interpret acoustic signals. In order to characterize pitch in a quantitative fashion, researchers have been using fundamental frequency (F0) as the physical surrogate of pitch, where F0 refers to the lowest frequency of a periodic sound waveform.
% Figure~\ref{fig:pitch_f0} shows the general relationship between pitch and F0.

It is widely recognized that the key predictor of the pitch of a sound is its \textbf{\textit{periodicity}}. The definition of F0 has an assumption that overtones are perfectly periodic in the frequency domain, and likewise acoustic signals are perfectly periodic in the time domain. However, this is not necessarily true in practice. For example, all string instruments produce overtones that are slightly sharper than ideal harmonics, due to effective shortening of the string at higher frequencies \cite{moorer1975segmentation, traube2000estimating}. The minor inharmonicity makes the sound waveforms not perfectly periodic. But since the deviation is only a few percent, the waveforms are called \emph{quasi-periodic}.

% \begin{figure}[t]
% 	\centering
% 	\includegraphics[width=0.8\linewidth]{Figures/pitch_f0.pdf}
% \caption{Relationship between pitch and F0.}
% \label{fig:pitch_f0}
% \end{figure}

So is human voice. From an articulatory point of view, F0 is the frequency at which vocal folds vibrate in voiced sounds. Voiced speech signals consist of periodic signals and random perturbations. The random component is mostly jitter and shimmer, which are quantified as the cycle-to-cycle variations of F0 and amplitude \cite{teixeira2013vocal}. Human voice is also deemed quasi-periodic with small perturbations.

Given the nature of musical sound and human voice, the abstract notion of F0 does not entirely apply to sound in the real world. Researchers in the past have investigated this phenomenon; e.g., \cite{shepard1964circularity} discusses \textit{Shepard tone illusion}, suggesting that perceived pitch can not be adequately represented by a purely rectilinear scale because of the circularity of pitch space \cite{burns1981circularity, braus1995retracing, deutsch2010paradox}. A substantial number of literature has discussed the impact of voice quality on facilitating pitch perception. \cite{swerts2001effect, honorof2005perception} is early work that presented the potential dependency of F0 on voice quality. And \cite{lee2009identifying, allen2014symmetric, kuang2015effect, kuang2015influence, kuang2016voice,  kuang2018integrating} further discover that changes in spectral shape affect people's judgements of relative pitch height. Besides spectral slope, \cite{ kuang2016effect} suggests that vocal fry, as another voice quality cue, would bias listeners to perceive a lower pitch. More generally, phonation is an important correlate of tone perception for various tonal languages \cite{brunelle2009tone, yang2011phonation, garellek2013voice, yu2014role}. There are abundant other non-f0 cues for pitch perception, including sound intensity \cite{neuhoff1996doppler} and temporal envelope \cite{kong2006temporal}, to name a few.

% Pitch is not F0 in production.
On the other hand, pitch does not exclusively depend on F0 in production. \cite{zhang2016mechanics} proposes that vocal folds tension plays an important role in the control of pitch in voice production. Similarly, \cite{titze1988framework} analyzes the correlation between different types of phonation and pitch production. Later, \cite{roubeau2009laryngeal, kuang2017covariation} show that voice quality is closely associated with pitch production in both speaking and singing voices. Specifically, lowest pitch production is associated with creaky voice, while the highest pitch production is associated with falsetto or whistle. \cite{rhee2019integration} also proves that voice quality helps the acquisition and production of Mandarin tones.

It is unsurprising that changes in pitch, as we will point out in this paper later, have covariates other than changes in the spacing of overtones, or the spacing of repetition in the time domain. Prior work has provided sufficient evidence, either implicitly or explicitly, of this argument.
% Prior work has indicated, either implicitly or explicitly, that in acoustic signals there exists internal spectral information that covaries with F0 as cues to pitch.
\cite{ryant2014highly, ryant2014mandarin} show that highly accurate Mandarin tone classification is possible using only MFCC features, when both F0 and overtone series are absent. \cite{lin2018improving} and \cite{zhang2018pitch} discuss the task of Mandarin tone classification and pitch range estimation respectively, but both work found that a model achieves better results with the addition of spectral features, compared to the one that uses prosodic features only.
% All these findings evidently show that covariates for pitch exist extensively in acoustic signals.

In order to better validate and understand the existence of latent information related to pitch in spectral features, we design experiments that track pitch on acoustic data where only F0 is varied. We study both synthetic and real human voice data. The goal is to see whether predicting F0 estimates from spectral features only is possible, and how difficult it would be. We believe that the results will suggest a promising direction for better modeling prosodic features in human voice.

\begin{table}[t]
 \caption{Statistics of pitch from human stimuli.}
 \label{tab:stimuli_stats}
 \centering
 \scalebox{0.85}{\begin{tabular}{c|c||c|c|c|c}
    \toprule
    \textbf{Subject} & \textbf{Property} & \textbf{max} & \textbf{min} & \textbf{avg} & \textbf{std}\\
    \midrule
    % \multicolumn{3}{c|}{\textbf{SAD}} 
    \multirow{4}{*}{\textbf{Female}} &  \textbf{\texttt{Period}} (s) & 1.17 & 0.40 & 0.5838 & 0.1605 \\\cline{2-6}
    & \textbf{\texttt{Pitch}} (Hz) & 747.07 & 67.87 & 300.80 & 113.37 \\\cline{2-6}
    & \textbf{\texttt{Duration}} (s) & 20.39 & 5.15 & 8.24 & 2.47 \\\cline{2-6}    
    & \textbf{\texttt{Amplitude}} (Hz) & 192.53 & 20.07 & 49.90 & 39.71 \\
    \midrule
    \hline
    \multirow{3}{*}{\textbf{Male}} &
        \textbf{\texttt{Period}} (s) & 1.32 & 0.40 & 0.7223 & 0.2375 \\\cline{2-6}
    & \textbf{\texttt{Pitch}} (Hz) & 363.02 & 64.37 & 149.38 & 43.39 \\\cline{2-6}
    & \textbf{\texttt{Duration}} (s) & 20.59 & 3.63 & 12.04 & 4.23 \\\cline{2-6}  
    & \textbf{\texttt{Amplitude}} (Hz) & 85.90 & 10.01 & 24.66 & 17.29 \\
    \bottomrule
 \end{tabular}}
\end{table}

%% file: 02-human.tex
\section{Human stimuli}
\label{sec:human_stimuli}

\begin{figure}[t]
	\centering
	\includegraphics[width=\linewidth]{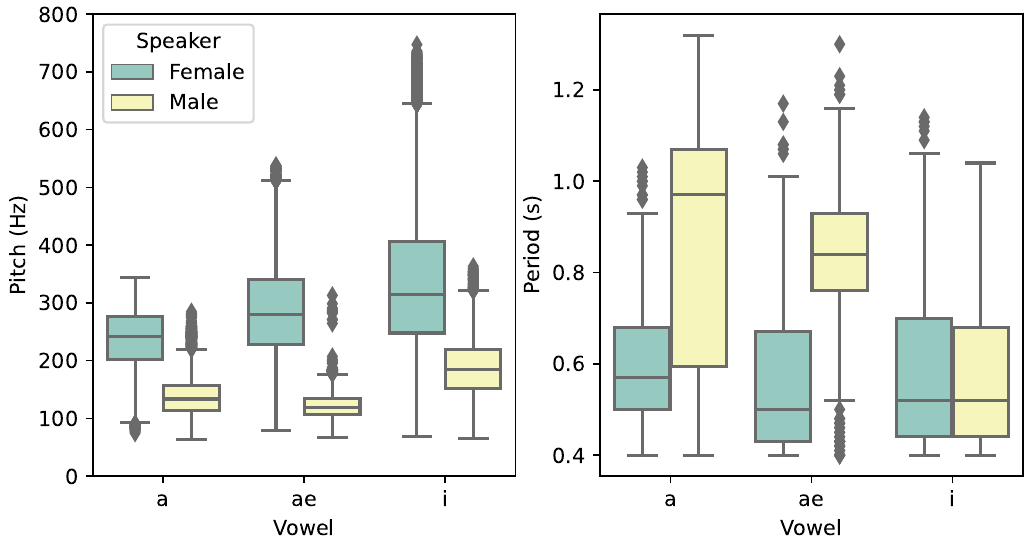}
\caption{Distribution of pitch range (\textit{left}) and period (\textit{right}) for each combination of speakers and vowels.}
\label{fig:pitch_period}
\end{figure}

\begin{figure*}[htp]
	\centering
    \subfigure[Train on one vowel one speaker]{
	\centering	\includegraphics[width=0.47\linewidth]{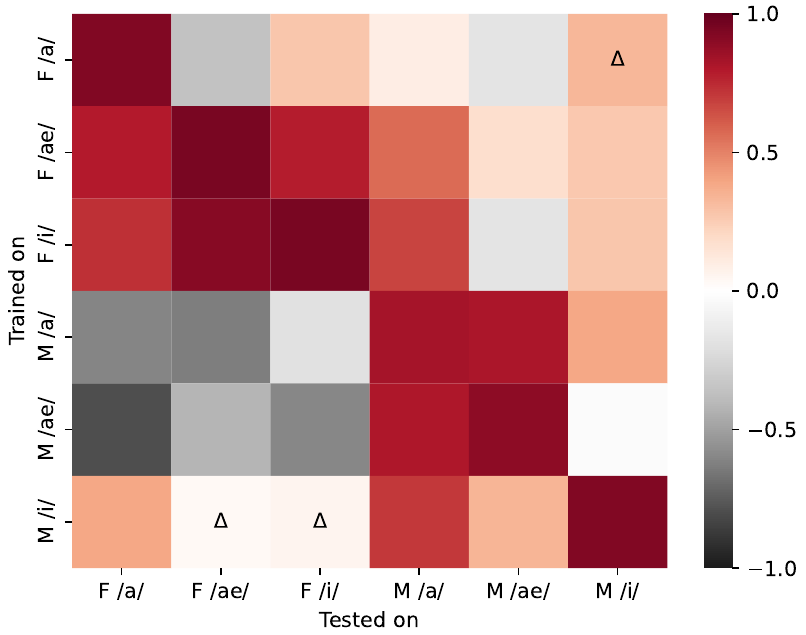}
	\label{fig:6x6_corr}}
	\hspace{1em}
    \subfigure[Train on all the vowel stimuli]{
	\centering	\includegraphics[width=0.47\linewidth]{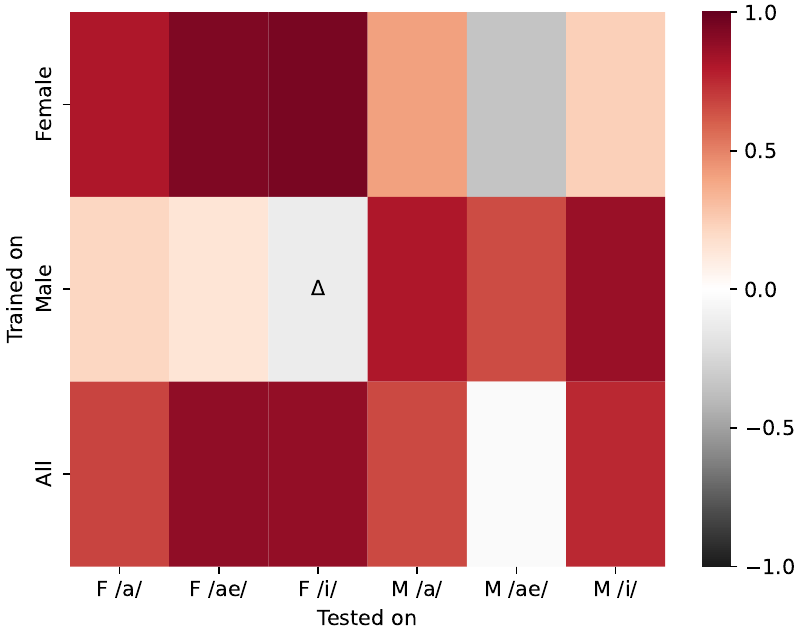}
	\label{fig:expand_corr}}
\caption{Pearson correlation coefficient between the predicted and gold semitones values for the experiments involving human data. Each cell represents one train/test set combination with the y-axis defining the training set and the x-axis the test set. Darker/lighter colors indicate stronger/weaker correlation. (a) Training on a single vowel from a single speaker. (b) Training on {\bf all} vowels from either a single or both speakers. {\bf Abbreviations}: F: female speaker,  M: male speaker, $\triangle$: result in cell is \emph{not} statistically significant (p $>$ 0.05)}
\label{fig:corr_cm}
\end{figure*}

\subsection{Data collection}
For the human experiments, we record two human subjects (one male and one female). They are instructed to sinusoidally vary pitch for three vowels: /\textipa{A}/, /\textipa{\ae}/ and /\textipa{i}/. For each vowel, subjects repeat this process multiple times until at least 5 minutes of usable utterances have been produced. Subjects are encouraged to vary the period, amplitude, and starting frequency of their oscillations from utterance to utterance, but to be careful to not vary these parameters within an utterance. Subjects are recruited from the student population at Anonymous University (age range: 20-30 years) and neither has professional voice training.

Table~\ref{tab:stimuli_stats} shows some major statistics of the collected stimuli, where \textbf{\texttt{Pitch}} is the F0 estimate acquired from RAPT algorithm \cite{talkin1995robust} as implemented in the Speech Signal Processing Toolkit (SPTK) \cite{sptk}. The following parameters are used:
\begin{itemize}
\itemsep0em
    \item \textbf{\texttt{wind\_dur}} = 10 ms
    \item \textbf{\texttt{min\_f0}} = 60 Hz
    \item \textbf{\texttt{max\_f0}} = 800 Hz for female, 400 Hz for male
\end{itemize}
% \texttt{wind\_dur}=10ms, \texttt{min\_f0}=60Hz, \texttt{max\_f0}=800Hz for the female speaker, and 400Hz for male.
\textbf{\texttt{Period}} is the period of sinusoidal oscillation across all the utterances, \textbf{\texttt{Duration}} is the duration of one utterance, and \textbf{\texttt{Amplitude}} is the difference between the highest and lowest pitch in each period. Figure~\ref{fig:pitch_period} gives a more detailed distribution of pitch values and period. As expected, the pitch range of the female speaker is more widely spread out; while the male speaker generally produces slower-varying vocalizations.

% \begin{table}[htb]
%  \caption{Amount of human stimuli.}
%  \label{tab:stimuli_amount}
%  \centering
%  \begin{tabular}{c||c|c|c}
%     \toprule
%     & \multicolumn{3}{c}{\textbf{Active speech} (mm:ss)}\\
%         \midrule
%     \textbf{Subject} & \textbf{/\textipa{A}/} & \textbf{/\textipa{\ae}/} & \textbf{/\textipa{i}/}\\\hline
%     \textbf{Female} & 5:01 & 5:02 & 5:01 \\\hline
%     \textbf{Male} & 5:04 & 5:06 & 5:03 \\
%     \bottomrule
%  \end{tabular}
% \end{table}

\begin{figure}[t]
	\centering
	\includegraphics[width=\linewidth]{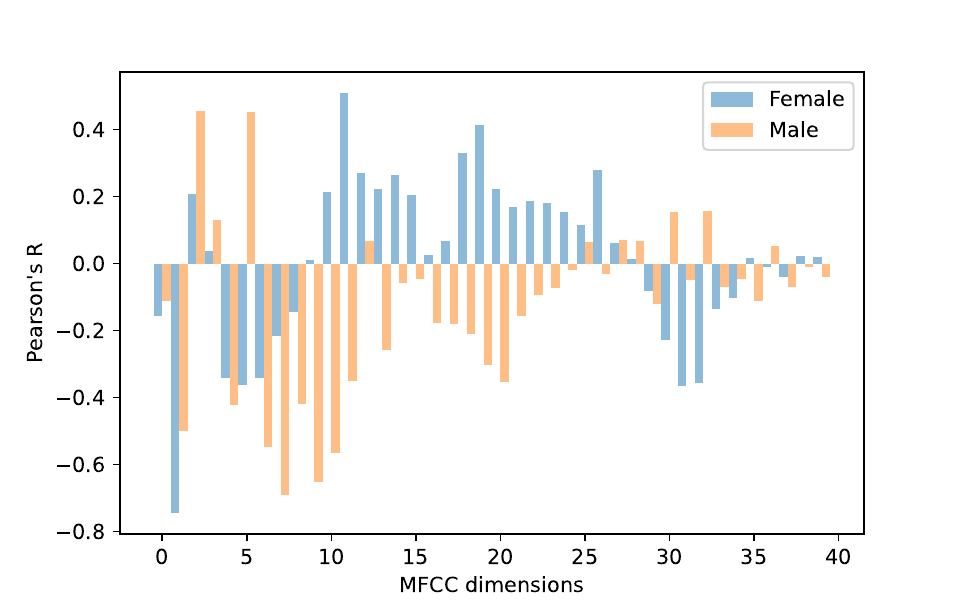}
\caption{Pearson correlation between F0 and each dimension of cepstral coefficients. The correlation is obtained from all the vowel stimuli collected from each speaker.}
\label{fig:both_corr}
\end{figure}

\subsection{Analysis}

% \begin{figure*}[htp]
%         %\begin{minipage}{0.35\textwidth}
%         % \hspace{-2.5em}
% 		\centering
% 		    \subfigure[Train and test on the same speaker.]{
% 			\centering
% 		\includegraphics[width=0.47\linewidth]{Figures/best_performance_semitone.pdf}
% 			\label{fig:best}}
% 		\hspace{1.5em}
%             \subfigure[Train and test on different speakers.]{
% 			\centering
% 			\includegraphics[width=0.47\linewidth]{Figures/worst_performance_semitone.pdf}
% 			\label{fig:worst}}
% 		\caption{Prediction of pitch of an excerpt of the vowel /\textipa{\ae}/ stimuli from the female speaker. (a) Predicted (pred) vs gold (RAPT) semitones for a sample utterance from female speaker. (b) Same as (a), except that the linear regression model is trained on the same vowel stimuli from the male speaker, and tested on female speaker.}
% 		\label{fig:less_train}
% \end{figure*}

Figure~\ref{fig:female_a_eg} plots the F0 estimates, spectrogram, and 40-D mel Frequency Cepstral Coefficients (MFCCs) (Section~\ref{sec:method} for details) corresponding to one of the female /\textipa{A}/ productions. The spectrogram is plotted using a 20 ms long Hanning window. Dynamic range is set to 45 dB, and pre-emphasis is 0 dB/Octave. For this production, the oscillation in F0 is clearly reflected in both the spectrogram and the MFCCs, particularly in the higher coefficients. Similar patterns are revealed in productions of the other vowels and for the male speaker.

Given the interesting patterns that emerged from visual inspection of MFCCs for isolated utterances, we compute the Pearson's correlation \cite{pearson1895notes} between F0 and each dimension of cepstral coefficients, plotted in Figure~\ref{fig:both_corr}. While all correlations are statistically significant, the correlations are particularly strong for 10th-28th cepstral coefficients. This pattern is somewhat stronger for the female than male speaker.

% To further understand the relationship between F0 and MFCCs, in Figure~\ref{fig:both_corr} we plot correlations between F0 and each cepstral coefficients, measured by Pearson's r \cite{pearson1895notes}. All the correlations are statistically significant.

% In accordance with Figure~\ref{fig:female_mfcc}, there is strong correlation between F0 and cepstral coefficients between the 10th to 28th dimensions for female data. Whereas that of male data is not salient. There are only sporadic, non-consecutive positive correlations appearing in the first few and last few dimensions.

% Gold F0 estimates are obtained from the pitch tracker in Praat \cite{boersma2011praat}. We then conduct two types of experiments, intra-speaker and inter-speaker pitch prediction. The acoustic features and model used are the same as described in Section~\ref{sec:synth_meth}.

%% file: 03-experiments.tex
\section{Experiments}
\label{sec:experiments}

\begin{figure*}[htp]
        %\begin{minipage}{0.35\textwidth}
        % \hspace{-2.5em}
		\centering
		    \subfigure[Train and test on the same speaker.]{
			\centering
		\includegraphics[scale=0.4]{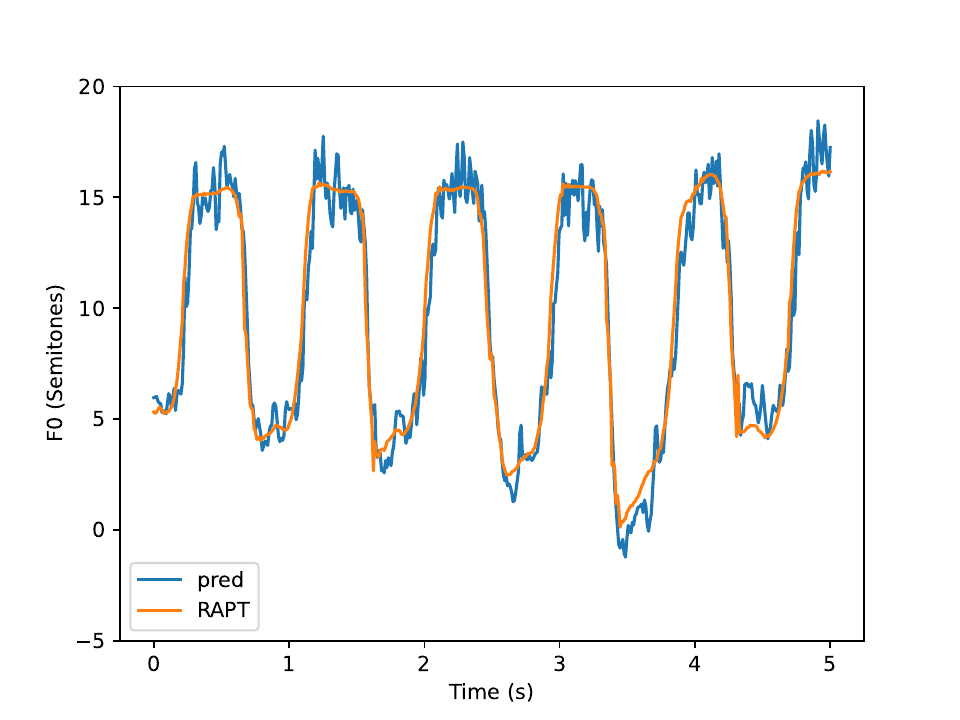}
			\label{fig:best}}
% 		\hspace{-1em}
        \subfigure[Different amounts of training data.]{
		\centering
    	\includegraphics[scale=0.36]{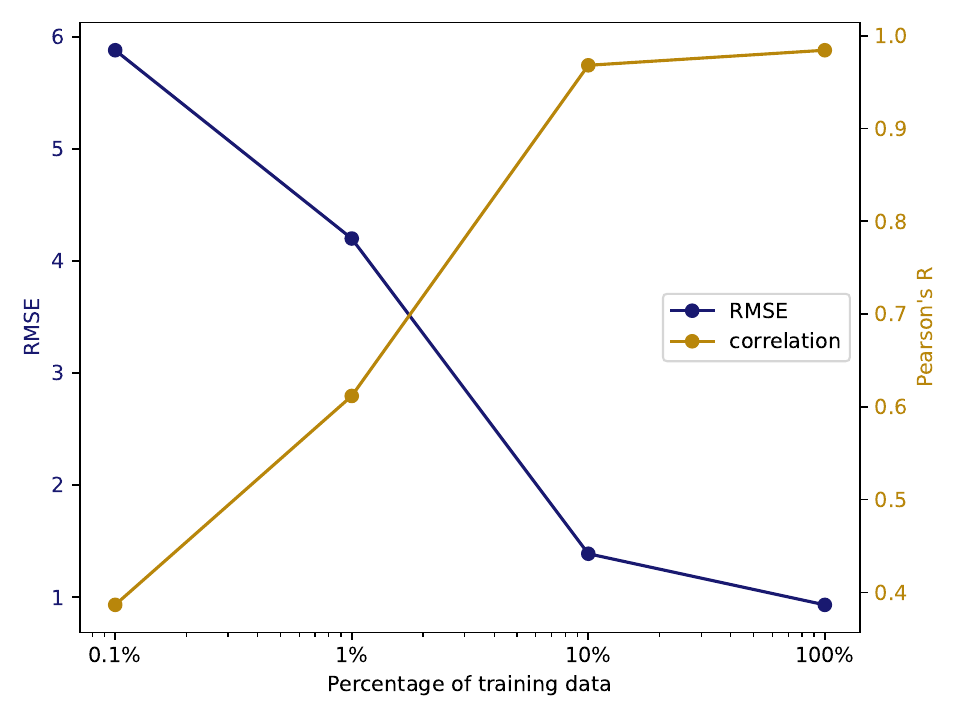}
			\label{fig:rmse_corr}}
        \hspace{-1em}
            \subfigure[Train and test on different speakers.]{
			\centering
			\includegraphics[scale=0.4]{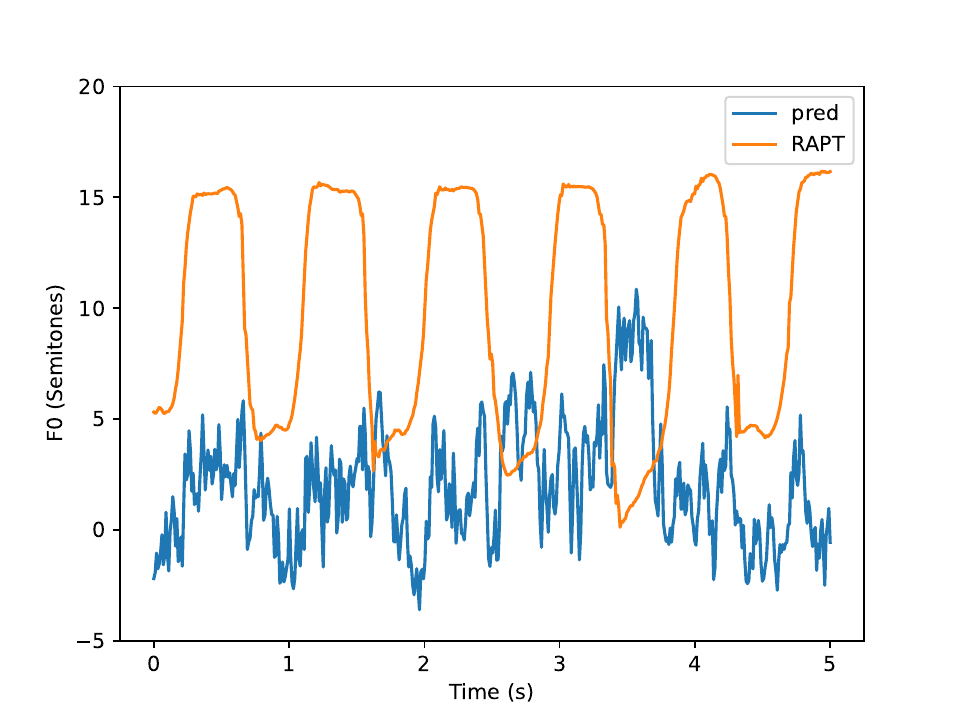}
			\label{fig:worst}}
		\caption{Prediction of pitch of an excerpt of the vowel /\textipa{\ae}/ stimuli from the female speaker. (a) Predicted (pred) vs gold (RAPT) semitones for a sample utterance from female speaker. (b) The RMSE and Pearson correlation between the gold pitch and predicted pitch based on the amount of training data. Full training data is \textbf{4 minutes}. (c) Same as (a), except that the linear regression model is trained on the same vowel stimuli from the male speaker, and tested on female speaker.}
		\label{fig:less_train}
\end{figure*}

\subsection{Method}
\label{sec:method}

\begin{itemize}[leftmargin=*]
    \item[] \textbf{Task} -- Each type of stimuli are partitioned into training and test set with an 80/20 split. The task is to predict time functions of F0. In order to alleviate the affect of different pitch ranges from different speakers (as shown in Fig~\ref{fig:pitch_period}), F0 values are transformed from Hz to semitones using the 5th percentile of each speaker as the base \cite{parish2016exploring}.
    We do a pairwise combination of training and test sets, whose results constitute a $6\times 6$ matrix shown in Figure~\ref{fig:6x6_corr}. The number that each cell represents is the average of 10 runs, and in each run the training/test split is different. This is to reduce the impact of randomness, as we have a fairly small dataset.
    % We will analyze the results in detail in Section~\ref{sec:discussion}.
    
    \item[] \textbf{Acoustic features} -- The collected recordings are resampled to 16kHz, and stored as mono 16-bit audio. Then we use 40-D Mel frequency cepstral coefficients (MFCCs) as the features of the input acoustic signals. MFCC coefficients are extracted using \textit{librosa} \cite{mcfee2015librosa} with a 10 ms step size and a 35 ms analysis window. % \texttt{TorchAudio} \cite{yang2021torchaudio} with a 10 ms step size.
    No overtone series is explicitly present in this representation.
    
    \item[] \textbf{Model} -- We utilize linear regression without using an L2 loss as implemented by the \textbf{\texttt{LinearRegression}} class in \textit{sklearn} \cite{pedregosa2011scikit} for all regression tasks.
    
    \item[] \textbf{Evaluation} -- Root-mean-square error (RMSE) and Pearson's R are used as evaluation metrics to measure the difference and correlation respectively between the ground truth and predicted semitone values.

\end{itemize}

% \begin{figure}[htb]
% 		\centering
%     	\includegraphics[width=\linewidth]{Figures/RMSE_corr_semitone.pdf}
% 		\label{fig:rmse_corr}\\[-1.5em]
% 		\caption{The RMSE and Pearson correlation between the gold pitch and predicted pitch based on the amount of training data. Full training data is \textbf{4 minutes}.}
% \end{figure}

\subsection{Results}

\begin{figure*}[htp]
		\centering
		  \subfigure[Sinusoidal stimuli of pitch]{
			\centering	\includegraphics[width=0.47\linewidth]{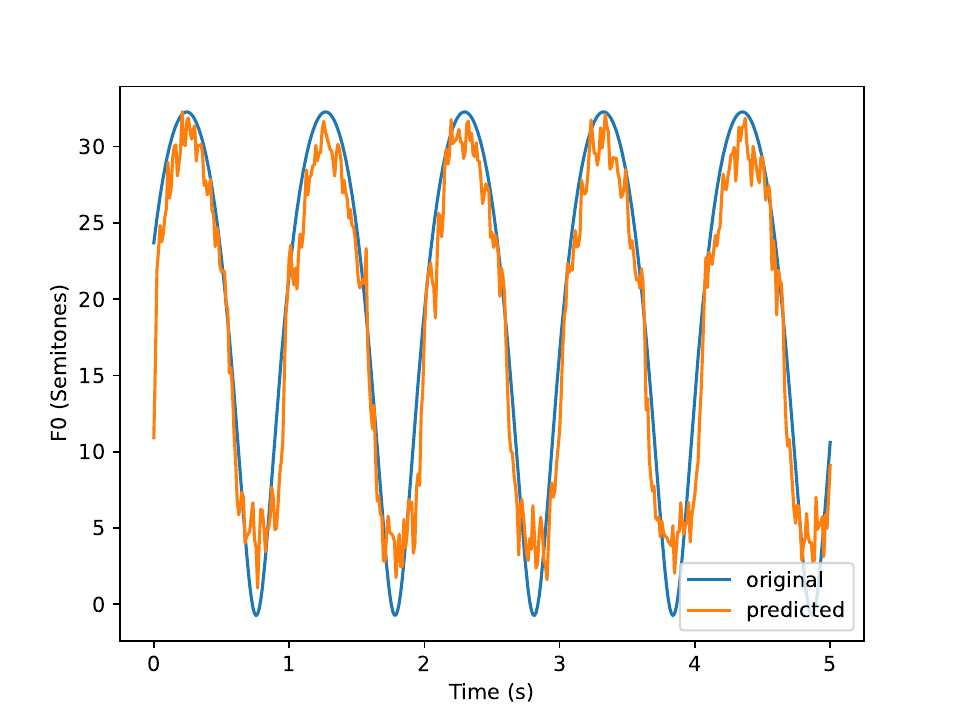}
			\label{fig:sine}}
		 \hspace{1.5em}
		  \subfigure[Complicated stimuli of pitch]{
			\centering	\includegraphics[width=0.47\linewidth]{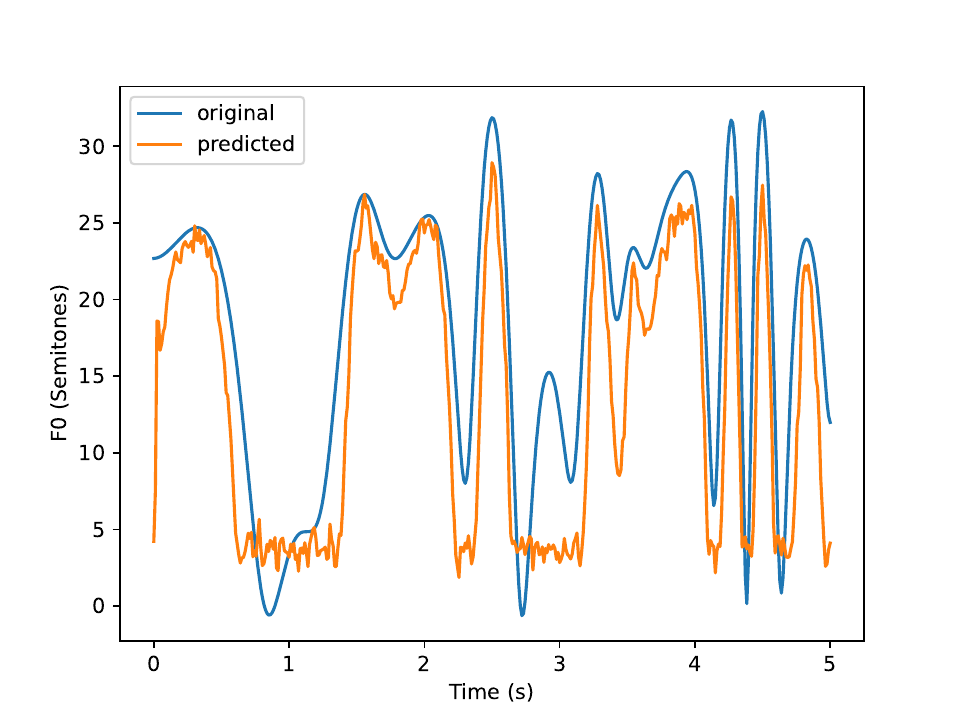}
			\label{fig:complicated}}\\[-1ex]
\caption{Synthetic stimuli from Pink Trombone and the predicted pitch.}
\label{fig:synthetic_stimuli}
\end{figure*}

Figure~\ref{fig:less_train} shows the most salient results of the experiments. When training and testing on the same vowel stimuli from the same speaker, even with such a small mount of training data and simplest regression model, MFCC spectral parameters are able to predict the pitch from real human voice. The predictions in Figure~\ref{fig:best} roughly fit the gold pitch tracks, except being a little jagged in peaks and troughs.

More interestingly, if the training data is reduced to 10\%, RMSE remains almost the same, meaning that around \textit{25 seconds} of MFCC features is enough for the model to predict the pitch well. More variations of training data and its corresponding performance can be found in Figure~\ref{fig:rmse_corr}. Knowing that RMSE is greatly affected by the pitch range of training data, we also plot Pearson's r between the gold pitch and prediction. The conclusion still holds. This is strong evidence of the argument that pitch has spectral covariates other than the overtone series.

But Figure~\ref{fig:worst} indicates that the covariates are not that simple. The linear regression model struggles to recover F0 if tested on different kinds of materials. Additional experiments show that by adding contexts to MFCC features and change to a more complicated model, such as multi-layer perceptron regressor, will improve the performance in both inter-speaker and cross-speaker settings. It is plausible that compared to linear models, MLP can better extract prosodic information from MFCC features.

%% file: 04-synthetic.tex
\section{Synthetic stimuli}
\label{sec:synth}

% \begin{figure}[ht!p]
%         %\begin{minipage}{0.35\textwidth}
%         \hspace{-2.5em}
% 		\centering
% 		    \subfigure[Train and test on the same speaker.]{
% 			\centering
% 		\includegraphics[scale=0.5]{Figures/best_performance_semitone.pdf}
% 			\label{fig:best}}\\
% % 		\hspace{-1em}
%         \subfigure[Different amounts of training data.]{
% 		\centering
%     	\includegraphics[scale=0.48]{Figures/RMSE_corr_semitone.pdf}
% 			\label{fig:rmse_corr}}\\
%         \hspace{-2.5em}
%             \subfigure[Train and test on different speakers.]{
% 			\centering
% 			\includegraphics[scale=0.5]{Figures/worst_performance_semitone.pdf}
% 			\label{fig:worst}}\\
% 		\caption{Prediction of pitch of an excerpt of the vowel /\textipa{\ae}/ stimuli from the female speaker. (a) Predicted (pred) vs gold (RAPT) semitones for a sample utterance from female speaker. (b) The RMSE and Pearson correlation between the gold pitch and predicted pitch based on the amount of training data. Full training data is \textbf{4 minutes}. (c) Same as (a), except that the linear regression model is trained on the same vowel stimuli from the male speaker, and tested on female speaker.}
% 		\label{fig:less_train}
% \end{figure}

We replicate the same task on synthetic data and it yields similar results. The conclusion is even stronger -- it is possible to recover much more complicated oscillations of pitch using the combination of single-frame MFCCs and a linear regression model. Below are the details of experiments and findings.

\subsection{Data generation}
%The F0 of glottal wave in Pink Trombone is varying within the range of [60, 404].

We generate synthetic data using Pink Trombone \cite{thapen2017pink}\footnote{Specifically, using the following Python re-implementation: \url{https://github.com/dkadish/pynktrombone}}, an articulatory synthesizer by varying the F0 of the glottal wave while keeping all other control parameters fixed. Five second duration utterances are generated by sampling the non-pitch control parameters, and the F0 contour is determined according to one of two mechanisms:

\begin{itemize}[leftmargin=*]
    \item[] \textbf{Sinusoidal stimuli.} F0 of the glottal wave is varied according to a sinusoid with the following form:
    \begin{equation}
      \text{F0} = 172 * \sin{(\alpha * t + \phi)} + 232
      \label{eqn:eq1}
    \end{equation}
    where $\alpha \in [1.7227, 8.6133]$ and $\phi \in [-\frac{\pi}{2}, \frac{\pi}{2}]$. The resulting F0 values lie in the range [60 Hz, 404 Hz].
    
    Both $\alpha$ and $\phi$ are randomly selected within respective ranges. Therefore, each of the stimuli has different amplitude, period and phase shift to increase the diversity of inputs.

    \item[] \textbf{Complicated stimuli.} F0 is generated as the superposition of two sinusoidal oscillations $A_t$ and $\omega_t$:
    \begin{equation}
      A_t = \sin{(\alpha_1 * t + \beta_1)}
      \label{eqn:eq3}
    \end{equation}
    \begin{equation}
      \omega_t = \cos{(\alpha_2 * t + \beta_2)}
      \label{eqn:eq4}
    \end{equation}
    \begin{equation}
      \text{F0} = 172 * A_t * \sin{(\omega_t * t + \phi)} + 232
      \label{eqn:eq5}
    \end{equation}
    where $\alpha_1, \alpha_2 \in [0.8613, 3.4453]$ and $\beta_1, \beta_2, \phi \in [-\frac{\pi}{2}, \frac{\pi}{2}]$.
\end{itemize}
%

% \subsubsection{Sinusoidal stimuli}
% \subsubsection{Complicated stimuli}
In order to simulate the settings in human voice, we generate 60 five-second long audio files for each kind of stimuli, resulting in five minutes worth of synthetic data for each mechanism. Then they are divided into training/test set with an 80/20 split as well. All the other experimental settings, including features, model, and the task, are the same as described in Section~\ref{sec:method}.

\subsection{Results}

Two samples of the regression results are shown in Figure~\ref{fig:synthetic_stimuli}. Results in Figure~\ref{fig:complicated} are surprisingly good. A simple linear regression model is able to mostly recover the intricate shape of the input.

In order to have a quantitative assessment of the performance, Table~\ref{tab:rmse_synth} lists the RMSE of the training and test set for the synthetic pitch prediction task. Much to our surprise, errors of different kinds of stimuli are close. It is reasonable to hypothesize that, regardless of the complexity of input signals, spectral features contain enough information to predict pitch.
% We defer more investigation of this hypothesis for future research.

\begin{table}[t]
  \caption{RMSE of synthetic pitch prediction.}
  \label{tab:rmse_synth}
  \centering
  \begin{tabular}{c||cc}
    \toprule
    \textbf{Stimuli} & \textbf{Training} & \textbf{Test}\\
    \midrule
    \textbf{\texttt{Sinusoidal}} & 3.079 & 2.733 \\\hline
    \textbf{\texttt{Complicated}} & 2.592 & 2.364 \\
    \bottomrule
  \end{tabular}
\end{table}

%% file: 05-discussion.tex
\section{Discussion}
\label{sec:discussion}

\subsection{Measure of pitch}
In this study, we use F0 as the ground truth of pitch, despite pointing out that F0 and pitch are not exactly the same. We argue that the claim should be orthogonal to how pitch is represented in the experiments. F0 is used because it is closely correlated with pitch, and it is also one of the cheapest measure of pitch to obtain. This work shows that F0 can be (mostly) recovered solely based on MFCCs, without the presence of overtone series. The results are sufficient enough to corroborate the covariation between spectral features and pitch.

\subsection{Complicated covariates}
The success of single-frame MFCC features predicting pitch is unexpected.
Experiments presented so far provide solid evidence of the existence of covariates.
% related to pitch in spectral features
But many aspects of the covariates remain obscure. We are particularly interested in the reason for performance degradation in Figure~\ref{fig:worst}.

We explore more diverse experimental settings in Figure~\ref{fig:corr_cm} to seek an answer.
% There exist various factors that affect the results.
Training and testing on vowels produced by the same speaker obviously work better than testing on vowels produced by a different speaker. But it is also observed that training on both speakers can mitigate disadvantages in cross-speaker testing.

Additionally, pairs of vowels that are located closer in IPA vowel chart, i.e., have similar articulatory features, yield better results when predicting pitch. For example, within the same speaker, models trained on /\textipa{A}/ have better performance when testing on /\textipa{\ae}/ than on /\textipa{i}/.

The failure in Figure~\ref{fig:worst} could be due to the lack of training data, but a more interesting hypothesis is not ruled out -- covariates are not simple in human speech. There might be higher-order interactions among properties of spectra regarding speakers, recording conditions, phonetic contexts etc. We defer more investigation of this hypothesis for future research.

\subsection{Better representations of pitch}
The existence of covariates also encourages researchers to find something other than F0 to describe pitch. An ideal representation of pitch would address current concerns of pitch tracking. It would not suffer from  frequent doubling/halving errors \cite{murray2001study}. And it should be homomorphic to human perception. Existing pitch tracking methods have the problem that a minuscule change in the input can cause substantial change in the output. The lack of continuity is generally not favorable for the interpretability of a representation.

Currently, the spiral properties of pitch and human auditory illusions introduce all the complication of representing pitch gracefully. In the future, we will work more in the direction of better modeling prosodic features.

%% file: 06-conclusion.tex
\section{Conclusion}
\label{sec:conclusion}

This paper shows a surprising phenomenon that coarse spectral representations are able to infer pitch without the presence of overtone series, by using even the simplest form of regression models. Therefore, the results present a strong support for the existence of covariates for pitch in coarse spectra. We validate the argument by experimenting on both human voice and synthetic articulations. And we further discover that, the covariates might be in a complex form and are affected by speakers and phonetic contexts. Altogether, this study provides a further understanding of prosodic features, and suggests a future research direction of better characterizing speech pitch.